\documentclass[11pt, oneside]{amsart}   	
\usepackage{geometry}                		
\usepackage{graphicx}				
\usepackage{amssymb}
\usepackage[hidelinks]{hyperref}
\hypersetup{
    colorlinks=false,
    pdfpagemode=FullScreen,
    }

\date{}
\title{Rod Burstall: In Memoriam}
\author{J Strother Moore}
\address{Department of Computer Science, University of Texas at Austin}
\email{moore@cs.utexas.edu}
\author{Gordon Plotkin}
\address{School of Informatics, University of Edinburgh}
\email{gdp@inf.ed.ac.uk}
\author{David Rydeheard}
\address{Department of Computer Science, University of Manchester}
\email{davidrydeheard@gmail.com}
\author{Don Sannella}
\address{School of Informatics, University of Edinburgh}
\email{dts@inf.ed.ac.uk}

\begin{document}

\maketitle
\vspace{-20pt}
\section*{Rod}

Rodney Martineau Burstall --- Rod, as he was known to us all --- died on Thursday, February 13th, 2025, after a long illness.
 Rod was a kind and generous man who will be remembered by those who knew him as much for his humanity as for
 his contributions to computer science.  While he made major contributions to our subject,
he
 constantly demonstrated humility, curiosity, openness, tolerance,
 and acceptance. 
 He read widely and enjoyed discussing --- or
 learning about --- basically any topic that his conversational partner felt
 passionate about.  Perhaps more than anything he exemplified a comfortable
 way of being human.  To many of us that was his greatest contribution to our
 lives.

Rod was born in 1934, the son of a draftsman and
a housewife from Liverpool.  He attended King George~V Grammar School
at Southport, moved on to King's College, Cambridge, reading Natural
Sciences, and then took a Masters in Operational Research at
Birmingham University. There followed a period in industry, working on
operational research at Bureau Gombert, Brussels and then the Reed
Paper Group in Kent. After the first year at Reed, he moved to the
programming group having written his first program while working in
Brussels.

In 1962 he returned to Birmingham, working as a research fellow at the
Department of Engineering Production and taking a Ph.D.\ on heuristic
programming for operational research 
 under the supervision of N.~A.\ Dudley and K.~B.\ Haley [3].
During this period --- the
early 1960s --- his interests began to broaden out into various
aspects of artificial intelligence and programming.  He had met Peter
Landin and then Christopher Strachey who introduced him to the then
very new mathematical ideas of programming language semantics; this
later became a major theme of his work at Edinburgh.

Rod's career at Edinburgh University, where he remained until his
retirement, began when Donald Michie invited him to join his
Experimental Programming Unit as a research fellow in 1964.  He became
a lecturer in the Department of Machine Intelligence and Perception in
1967, a Reader in the School of Artificial Intelligence in 1970, and
was promoted to a personal chair in Artificial Intelligence in the
department of the same name in 1977. In 1979 he moved from the
Department of Artificial Intelligence to the Department of Computer
Science, joining with Robin Milner, Gordon Plotkin and others to form
a very strong group in theoretical computer science.  This happened at
a time when the theory subdiscipline was becoming more clearly part of
computer science and somewhat detached from artificial intelligence.
In 1980 his personal chair was retitled ``Computer Science''. He retired 
from the university in April 2000, as Professor Emeritus, but continued to visit the university from time to time.

Rod maintained an impressive group of post-docs, Ph.D.\ students and
visitors, partly supported by a continuous stream of funded projects.
He  had visiting positions in the USA, France, Belgium and Japan,
and, among others, industrial consultancies at Xerox Parc, IBM
Laboratories UK, ICL Kidsgrove and Digital Equipment Corporation's
System Research Center. He was an invited speaker at a range of
conferences in Europe, the United States and Japan. He was elected to
the Academia Europea in 1989 and the Royal Society of Edinburgh in
1995. In 2009, he was awarded the Association for Computing Machinery 
(ACM) SIGPLAN Programming Language Achievement Award.

Much of Rod's most important contributions were informal; he was
always  concerned with building up community, whether among his
research team, or more widely within the Department of Computer
Science and, latterly, the School of Informatics. One manifestation
of this commitment was his starting an informal weekly evening seminar in 1973,
rotating between Rod and Robin Milner's homes.
Rod called it the BBMS (the 
Bob Boyer Memorial Society): Bob had not passed away, he had 
just gone back to Texas, but he had made the initial suggestion. This eventually led  
to the formation of the present-day Laboratory for
Foundations of Computer Science at Edinburgh in 1986 by Rod,
Matthew Hennessy, Robin Milner, and Gordon Plotkin. He served as
its director for three years.

Rod was sustained and supported by his Finnish wife Sissi, whom he
married in 1959. They had three daughters, Kaija, Taru, and Taina.  Together they created a
warm, welcoming and loving family atmosphere from which generations of his
graduate students and post-docs benefited. Sadly, Sissi died in 1990,
and his daughter Kaija died in 1994.   Rod was fortunate to have two 
granddaughters, Larissa and Alexandria. For the last 25 years of his life, Rod and Laila Kjellstr\"{o}m were  
partners and loving companions. 

As well as his academic life, Rod had a long-standing interest in Buddhism
both as a student and as a practitioner. He took temporary vows at one stage, and
happily combined the life of a university professor with wearing
Buddhist robes.

Rod was supported throughout his professional life by his personal 
secretary, Eleanor Kerse.  Eleanor worked with Rod from the mid-1960s until his
retirement, protecting him from the
demands of the outside world, and striving to keep him organized (the last a futile enterprise, her
superhuman efforts notwithstanding).


\section*{Rod's Scientific Achievements}

We try to summarize Rod's 
scientific achievements.  In addition, we describe the 
personal style and enthusiasm that he  brought to the subject. 
  In the development of our understanding of computing, Rod's work was
truly pioneering and seminal.  Not only did he contribute to
the early development of a wide range of topics,  he also
branched out into new and often quite unorthodox territories with
ideas that were striking in their originality. He  never ploughed a
single furrow for long, rather taking a new technical direction every 
few years and leaving the further development to others.  Some of these
forays developed into what are now large areas of research. There are perhaps
other pioneering ideas in his work which are yet to develop into scientific
maturity.

Rod's research was not only founded in practice --- in the actual
practice of writing programs, designing languages, and doing proofs
 --- but  at a point where theory and practice often cannot and
should not be distinguished. This close interplay was  a constant
characteristic of his work and often resulted in a very
distinctive approach to research topics.

Rod's idea of ``doing research'' was a delight: a very human occupation
--- real fun and a form of intellectual play. No-one who was
closely involved in research with Rod can forget the experience, and the
enthusiasm and pleasure that he brought to the activity.  Also
important was the sense that research is a community activity, shown in
his fostering (sometimes almost father-like) of his research group,
and in his concern about the wider community of researchers.

Trying to summarise Rod's scientific contribution is not easy.
His interests ranged widely, and yet the interplay of constant
themes and aims is always apparent.  Indeed, Rod himself may well not
have appreciated the notion of trying to ``classify'' his work into areas
when it is so obviously an organic whole. However, in order to
describe the range of his activity we will necessarily have to lapse
into describing topics and areas of development.



Rod's early work and his Ph.D.\ topic was in Operational Research, in
particular developing heuristics for solving specific problems,
and implementing
these as computer programs. These early ideas are reflected in his two
major areas of research: Artificial Intelligence and what one could
broadly describe as Software Development.

Much of Rod's early work at Edinburgh was in Artificial Intelligence
in a group headed by Donald Michie who had worked with Turing during
World War~II at Bletchley Park.
This was an exciting era in the subject with a host of ideas, systems, 
and languages being developed, and a fruitful interplay with areas 
such as linguistics and psychology. One of Rod's major contributions
during this period was leading the team that programmed Freddy, the
first hand-eye assembly robot. As part of this he developed an early
vision recognition program which computed partial isomorphisms between
relational structures, and also contributed to natural language
processing.
The style of his work at this time is expressed well by the
following quote from his paper [\ref{machine-intelligence-survey}] in 1970:
\begin{quote}
  ``we are coming to a better understanding of the algebraic and
  logical principles underlying the problems which have been attempted
  so far.  I am thinking here of the use of first order logic in
  question answering and problem solving programs, of a more rigorous
  set-theoretic approach in describing puzzle solving and game playing
  programs, [and] of studying inductive generalisation in the context of
  logic and automata theory''
\end{quote}
Much of his effort during this period was devoted to the development
of the POP-2 programming language (see below), which was directly based
on emerging ideas about underlying principles of programming languages
and was used for all AI programming in Edinburgh at that time.

During this period, Rod became interested in a mathematical approach to
software development. In those days (the mid-1960s) this was a radical
idea and the links between mathematics and program design were
recognised by very few.  Amongst them were Peter Landin and
Christopher Strachey. Rod met Peter Landin through chance acquaintance
in a London bookshop,
and was later introduced to Christopher Strachey who was Landin's
employer at the time.  Their informal and almost clandestine
meetings and their subsequent influence on Rod's work are engagingly
described in [\ref{strachey}].
Through these meetings, Rod
was introduced to new and exciting ideas, including universal algebra
and early ideas about functional programming,
which led him to a series of
fundamental innovations in the development and application of
mathematics to computer programming.  If one were to summarise the main
theme of this work, it lies in the dream that we ought to be able to
say in a mathematical language what a program is supposed to do and
then prove that it does indeed do this. Behind this simple aim,
however, lies a formidable requirement: to develop new mathematics
--- especially in mathematical logic, abstract algebra and topology
--- and to use this to invent new approaches to programming and to
understanding programming languages.

This then sets the scene for Rod's major contributions. Rod was among
the first to study the problem of proving properties of programs.  He
was the earliest to recognize
the central role of structural induction in establishing properties of
data and programs, giving a clear and appealing exposition of this
fundamental technique in [\ref{structural-induction}]. This paper also 
introduced the idea of defining recursive functions on datatypes by 
pattern-matching on their  constructors.
While not named as such in the paper, this  
is structural recursion, the natural programming companion to structural induction
(cf.\, induction on natural numbers and primitive recursion).
Rod's work on reasoning about
mutable data structures [\ref{mutable}] was an early and influential precursor 
of separation logic, as later developed  by John C.\ Reynolds, 
Peter O'Hearn, Samin Ishtiaq and Hongseok Yang; in particular,
the basic idea of spatial conjunction is implicit in his work.

In [\ref{intermittent-assertions}] Rod developed the so-called ``intermittent assertion'' method of
program proving, in which an
assertion attached to a point in a program means that control will
\emph{sometime} pass that point and satisfy the assertion, rather than
satisfying the assertion \emph{whenever} it (perhaps never) passes
that point, as advocated by Bob Floyd, Tony Hoare and Peter Naur; this allows total
correctness to be established with a single proof.
This paper was the first
to point out the connection between program proof and modal logic, an
idea that was developed by Amir Pnueli and many others into what is
now a major research area, with particular applications to specifying
and reasoning about concurrent systems.

To reason about what programs do, it is necessary to formalise 
the semantics of the programming language. The mathematical foundations of 
this area were being developed from the mid-1960s, in particular 
through the work of Dana Scott and Christopher Strachey. Rod was active in 
the area, investigating the semantics of languages with a notion of
``state'' in [\ref{assignment}] and showing how to define programming
language semantics in first-order logic in [\ref{first-order}].
Rod was responsible for the idea, used by Strachey and
thereafter, that the store is a mapping from locations (``L-values'')
to their contents.  His student Mike Gordon wrote a denotational
semantics for LISP, and Rod encouraged his post-doc Gordon Plotkin 
in his early work on denotational semantics.

As part of his quest for correctness of programs, Rod, together with
his student John Darlington, undertook the first major work on program
transformation, whereby programs whose correctness has been
established are transformed into equivalent and possibly more
efficient programs through correctness preserving transformations~[\ref{darlington}].
They isolated a range of these transformations and implemented a
partly automated system. This inspired a fruitful worldwide research
movement, being viewed as a possible way forward in the development of
methods for creating more reliable, provably correct, programs.

One of the major themes supporting his work on program correctness was
Rod's long-standing interest in novel programming languages. This
began with the POP-2 language developed with Robin Popplestone and
others in the 1960s, which was a blend of LISP and Algol strongly
influenced by Strachey's CPL and early functional programming ideas of
Landin: see, e.g.,~[\ref{Pop2}].
 Later, Rod devised experimental functional programming languages, beginning with NPL~[\ref{NPL}], 
 which grew out of his work with John Darlington on program transformation.
 NPL evolved into Hope 
(nothing to do with faith and charity: at the
time, his research group was located in a
warren of small Edinburgh houses
in Hope Park Square, named after Thomas Hope, a 18th-century
Scottish agricultural reformer).  Hope~[\ref{Hope}] was a small
language adapted to correctness proofs and transformations, but it had
some useful and novel features, including the elegant and now-familiar
combination of algebraic datatypes
\begin{quote}
\begin{verbatim}
data tree(alpha) == empty ++ node(tree(alpha),alpha,tree(alpha))
\end{verbatim}
\end{quote}
with pattern-matching clausal function definitions
\begin{quote}
\begin{verbatim}
dec flatten: tree(alpha) -> list(alpha)
--- flatten(empty) <= []
--- flatten(node(t1,n,t2)) <= flatten(t1)<>(n::flatten(t2))
\end{verbatim}
\end{quote}
as well as a simple module system.
The language Standard~ML and its close relatives Caml and OCaml have become
the most widely used of the so-called ``strict'' or ``eager''
functional languages. They
combine features from Hope with those of their precursor, ML, devised
by Robin Milner; Rod was an active member of the Standard~ML
design team.  Another theme in Rod's work is the role of
abstraction in programming and related issues around modularity of
code.  Standard~ML incorporates a sophisticated system for
describing program modules, devised by Dave MacQueen, who worked on
Hope with Rod as a post-doc, and was also influenced by Rod's work on
modular specifications with Joseph Goguen (see Clear below).  Another language that Rod devised,
with Butler Lampson, was Pebble (``small and hard''!) that moved
programming languages towards type theories, using dependent types
for the interfaces of modules.  Pebble~[\ref{pebble}] was an
experimental language based on the $\lambda$-calculus
which, using only a few constructs, could express a
wide variety of programming language features: modules,
interfaces and implementations, abstract data types, generic types,
recursive types and unions.

Logic clearly has a role in program correctness, but so too does
algebra, and Rod pioneered the use of algebraic techniques in
programming. With Peter Landin, he used algebraic ideas --- algebras,
homomorphisms, commuting diagrams --- to structure proofs of correctness,
showing how to use this for the correctness of compilers. The role of
category theory is important here as a tool for describing structural
aspects at the right level of abstraction. Again Rod pioneered this
fusion of two such apparently disparate areas.
On one hand, he was an early user of categorical ideas in computing,
for example in describing programs as free categories, and interpreting
recursive program schemes, and later in devising ``specification
languages''; on the other, he saw that category theory itself
could be implemented as programs.  The latter led to the development of
Computational Category Theory [\ref{CCT}] together with Rod's student David
Rydeheard where constructions in category theory are expressed as
programs, producing code of unusual abstract functionality.
Nowadays, category theory is routinely used to formulate ideas in computing, 
in particular in the mathematical semantics of programming languages, 
but in those early days of the
1970s, it was viewed 
as a very exotic branch of mathematics and its application to computing was 
a ``leap of faith''.

The development of algebraic specification techniques owes much to
Rod's long-standing collaboration with Joseph Goguen who had done some
of the first work in this area as a member of the so-called ADJ group
at IBM.
They proposed the first algebraic specification language,
Clear~[\ref{clear-theories}], which focussed on the provision of mechanisms for combining
specifications of program behaviour.
(Many colleagues found the semantics of Clear~[\ref{clear}], expressed in unfamiliar
categorical language, complicated and difficult to understand.  In
reaction, Jacques Loeckx in Saarbr\"ucken later invented a specification
language called Obscure.)
Then, in trying to describe the semantics
at the right level of abstraction,
they introduced a notion of abstract logical system which they called
``institutions'' [\ref{intro-institutions}] and [\ref{institutions}].
The aim was to describe how specification modules could be combined or
parametrised independently of the nature of the individual modules.
This work, and their ideas in [\ref{CAT}] about ``vertical composability'' of
refinement steps (i.e., the correctness of \emph{stepwise} refinement)
and ``horizontal composability'' (i.e., compatibility of refinement
with specification structure), inspired by two-dimensional categories,
have had great influence.
This general line of work has been intensively
developed by a worldwide community of researchers including Rod's
student Don Sannella and post-doc Andrzej Tarlecki.

The interplay between the activities of programming and proving
programs correct, and actual code, is again evident in Rod's contributions
to the development of automated proof support systems.
His work on this began in the late 1960s with
[\ref{structural-induction}] and [\ref{first-order}] and continued in the
early and mid 1970s with theses on the topic by his students J Strother Moore, Rodney
Topor, and Raymond Aubin.  Later, he developed IPE (``Interactive Proof
Editor'') in the mid-1980s, and in
the 1990s led Randy Pollack, Zhaohui Luo and a team of other students
and post-docs in the development and use of the Lego proof assistant~[\ref{Lego}].
Lego implements a number of related type systems including the
Edinburgh Logical Framework, Coquand's Calculus of Constructions, and
Luo's Unified Theory of Dependent Types, supporting interactive proof
development in the natural deduction style and generating explicit
proof objects.  In this framework, Rod with his student James McKinna
investigated notions of programs packaged with proofs of their
correctness, introducing the idea of ``deliverables'' [\ref{deliverables}], a precursor to
the 
topic of ``proof-carrying code''.
He returned to the topic of making proof fun and easy for
novices with his ProveEasy system~[\ref{PE}] which he used in teaching logic to
undergraduates in the late 1990s.

Much of Rod's work is collaborative with other research workers and
with his research students and post-docs, often with long-standing
collaborations.  Unfortunately, in order to keep this summary to a
reasonable length, we have not been able to mention all of the many
collaborators in his work.  The list of co-authors in the bibliography
below documents some of these collaborations.

Influenced by his Buddhism, Rod became
interested in aspects of consciousness and our perception of reality
in relation to our experience of computing.
 We end
this summary of Rod's  work with a quote from one of his papers on this
subject [\ref{reality}]:
\begin{quote}
  ``In working with a computer we interact with a small world, partly of
  our own creation, in which we have a special role. We are the agent
  who makes things happen. In this world we play the role
  traditionally assigned to God. We have complete power and our aim is
  to control everything that happens in this world. The better we are
  at programming the more nearly we approach total control over what
  happens \ldots{}  Contrast this with an attitude of respect for other
  inhabitants of our world, other people, animals or forests, a view
  of the world in which we do not have some distinguished role. In
  such an attitude we are open to the richness of phenomena over which
  we have no dominion. Think of sitting on a hillside watching clouds
  move through the sky, as opposed to sitting at your terminal
  \ldots{} \\
  
  Computing carries a great deal of energy in our current culture, and
  fuels our curiosity and inventiveness. But in order to fully enjoy
  its possibilities we need to appreciate the way it can subtly
  influence our frame of thought. The recognition of this influence
  does not free us; but it may provide a starting point for us to look
  at ways of working with computers without being entrapped by a
  limited perspective based on our desire for control and exclusive
  reliance on conceptual thought.''
\end{quote}


\bigskip



\section*{Rod's students}

\begin{enumerate}

\item Gordon D.\ Plotkin, 1972, Automatic methods of inductive inference.

\item  John Darlington, 1973, A Semantic Approach to Automatic Program Development.

\item  Michael J.~C.\ Gordon, 1973, Evaluation and Denotation of Pure LISP Programs: a worked example in semantics.

\item  J Strother Moore, 1973, Computational Logic: Structure sharing and proof of program properties.

\item  Rodney Topor, 1975, Interactive program verification using virtual programs, 1975.

\item  Raymond Aubin, 1976, Mechanizing structural induction. 

\item  Martin Stephen Feather, 1979,  A system for developing programs by transformation.﻿

\item  David  Eric Rydeheard, 1981, Applications of Category Theory to Programming and Program Specification.

\item  Alan Mycroft, 1982,	Abstract interpretation and optimising transformations for applicative programs.﻿ ﻿

\item  Donald Sannella, 1982, Semantics, implementation and pragmatics of Clear, a program specification language.

\item  Alberto Pettorossi, 1984, Methodologies for Transformations and Memoing in Applicative Languages.


\item  Tatsuya Hagino, 1987, A categorical programming language.

\item  Oliver Schoett, 1987, Data abstraction and the correctness of modular programming. ﻿

\item  Martin Illsley, 1988, Transforming Imperative Programs.

\item  Brian Ritchie, 1988, The Design and Implementation of an Interactive Proof Editor.

\item  Zhaohui Luo, 1990, An extended calculus of constructions.﻿

 \item  James  H.\ McKinna, 1992, Deliverables: a categorical approach to program development in type theory.
 
\item  Michael  V.\ Mendler, 1992, Modal logic for handling behavioural constraints in formal hardware verification. ﻿

\item  Thorsten	Altenkirch, 1993, Constructions, inductive types and strong normalization. ﻿

\item  Healfdene Goguen,	1994, A Typed Operational Semantics for Type Theory. ﻿

\item  Robert Pollack, 1994, The Theory of LEGO: A Proof Checker for the Extended Calculus of Constructions.

\item  Makoto Takeyama, 1995, Universal Structure and a Categorical Framework for Type Theory.

\item Masahito Hasegawa, 1997, Models of sharing graphs: a categorical semantics of let and letrec.

\item  Thomas Kleymann, 1998, Hoare Logic and VDM: Machine-Checked Soundness and Completeness Proofs.

\item  Conor McBride, 1999, Dependently Typed Functional Programs and their Proofs.\\

\end{enumerate}

\subsection*{Bibliography}

\begin{enumerate}

\item
R.M. Burstall, R.A. Leaver and J.E. Sussams.
Evaluation of transport costs for alternative factory sites ---
a case study.
\emph{Operational Research Quarterly} 13:345--354 (1962).

\item
R.M. Burstall and G.R. Kiss.
Information processing language~V for the KDF-9 computer.
\emph{Automatic Programming Information} 18 (1963).

\item
R.M. Burstall.
Heuristic and decision tree methods on computers: some
operational research applications.
 Ph.D.\ thesis, University of Birmingham (1966).

\item
R.M. Burstall.
Computer design of electricity supply networks by a
heuristic method.
\emph{Computer Journal} 9(3):263--274 (1966).

\item
R.M. Burstall and J.V. Oldfield.
A software stack system for the 340/347 display.
\emph{Decuscope} 5(3):1--3 (1966).

\item
R.M. Burstall.
A heuristic method for a job-scheduling problem.
\emph{Operational Research Quarterly} 17:291--304 (1966).

\item
R.M. Burstall.
Tree searching methods with an application to a network
design problem.
\emph{Machine Intelligence 1} (eds.\ N.L. Collins and D. Michie).
Edinburgh: Oliver and Boyd, 65--85 (1967).

\item \label{assignment}
R.M. Burstall.
Semantics of assignment.
\emph{Machine Intelligence 2} (eds.\ E. Dale and D. Michie).
Edinburgh: Oliver and Boyd, 3--20 (1968).

\item \label{pop2}
R.M. Burstall, J.S. Collins and R.J. Popplestone.
\emph{POP-2 Papers.}
Edinburgh: Oliver and Boyd (1968).

\item
R.M. Burstall and R.J. Popplestone.
POP-2 reference manual.
\emph{Machine Intelligence 2} (eds.\ E. Dale and D. Michie).
Edinburgh: Oliver and Boyd, 205--246 (1968).
Also in [\ref{pop2}].

\item
D. Michie, A. Ortony and R.M. Burstall.
\emph{Computer programming for schools:  first steps in Algol}.
Edinburgh: Oliver and Boyd (1968).

\item
R.M. Burstall and J.S. Collins.
An introduction to the POP-2 programming language.
In [\ref{pop2}] (1968).

\item
J.S. Collins, A.P. Ambler, R.M. Burstall, R.D. Dunn, D. Michie,
D.J.S. Pullin and R.J. Popplestone.
Multi-POP/4120: a cheap on-line system for numerical and non-numerical
computing.
\emph{Computer Bulletin} 12(5):186--189 (1968).

\item
R.M. Burstall.
Writing search algorithms in functional form.
\emph{Machine Intelligence 3} (ed.\ D. Michie).
Edinburgh University Press, 373--385 (1968).

\item \label{structural-induction}
R.M. Burstall.
Proving properties of programs by structural induction.
\emph{Computer Journal} 12(1):41--48 (1969).

\item
R.M. Burstall.
A program for solving word sum puzzles.
\emph{Computer Journal} 12(1):48--51 (1969).

\item \label{landin}
R.M. Burstall and P.J. Landin.
Programs and their proofs:  an algebraic approach.
\emph{Machine Intelligence 4} (eds.\ B. Meltzer and D. Michie).
Edinburgh University Press, 17--43 (1969).


\item \label{first-order}
R.M. Burstall.
Formal description of program structure and semantics in first
order logic.
\emph{Machine Intelligence 5} (eds.\ B. Meltzer and D. Michie).
Edinburgh University Press, 79--98 (1969).

\item \label{machine-intelligence-survey}
R.M. Burstall.
Machine intelligence research at Edinburgh University.
\emph{Proc.\ ACM Intl.\ Computing Symposium}, Bonn, 696--708 (1970).

\item \label{Pop2}
R.M. Burstall, J.S. Collins and R.J. Popplestone.
\emph{Programming in POP-2.}
Edinburgh University Press (1971).
A revision of [\ref{pop2}] incorporating much new material.

\item
H.G. Barrow, A.P. Ambler and R.M. Burstall.
Some techniques for recognising structures in pictures.
\emph{Proc.\ Intl.\ Conf.\ on Frontiers of Pattern
  Recognition}, Honolulu.
Academic Press, 1--29 (1972).

\item
R.M. Burstall.
An algebraic description of programs with assertions,
verification and simulation.
\emph{Proc.\ ACM Conf.\ on Proving Assertions About
  Programs}, Las Cruces.
\emph{SIGPLAN Notices} 7(1):7--14 (1972).


\item \label{mutable}
R.M. Burstall.
Some techniques for proving correctness of programs which
alter data structures.
\emph{Machine Intelligence 7} (eds.\ B. Meltzer and D. Michie).
Edinburgh University Press, 23--50 (1972).

\item
D. Michie, A.P. Ambler, H.G. Barrow, R.M. Burstall, R.J. Popplestone
and K.J. Turner.
Vision and manipulation as a programming problem.
\emph{Proc.\ 1st Conf.\ on Industrial Robot Technology},
Nottingham, 185--190 (1973).

\item \label{freddy}
A.P. Ambler, H.G. Barrow, C.M. Brown, R.M. Burstall and R.J. Popplestone.
A versatile computer-controlled assembly system.
\emph{Proc.\ 3rd Intl.\ Joint Conf.\ on Artificial
  Intelligence}, Stanford, 298--307 (1973).

\item \label{darlington}
J. Darlington and R.M. Burstall.
A system which automatically improves programs.
\emph{Proc.\ 3rd Intl.\ Joint Conf.\ on Artificial
  Intelligence}, Stanford, 479--485 (1973).

\item \label{intermittent-assertions}
R.M. Burstall.
Program proving as hand simulation with a little induction.
Invited paper, \emph{Proc.\ IFIP Congress '74}, Stockholm,
  308--312 (1974).

\item
R.M. Burstall and J.W. Thatcher.
The algebraic theory of recursive program schemes.
\emph{Proc.\ 1st Intl.\ Symp.\ on Category Theory Applied to
  Computation and Control}, San Francisco, 1974.
Springer Lecture Notes in Computer Science, Vol.\ 25,
126--131 (1975).

\item \label{reliable}
R.M. Burstall and J. Darlington.
Some transformations for developing recursive programs.
Invited paper, \emph{Proc.\ Intl.\ Conf.\ on Reliable
  Software}, Los Angeles, 465--472 (1975).

\item
A.P. Ambler, H.G. Barrow, C.M. Brown, R.M. Burstall and R.J. Popplestone.
A versatile system for computer-controlled assembly.
\emph{Artificial Intelligence} 6(2):129--156 (1975).
Based on [\ref{freddy}].

\item
J. Darlington and R.M. Burstall.
A system which automatically improves programs.
\emph{Acta Informatica} 6:41--60 (1976).
Revised version of [\ref{darlington}].

\item
H.G. Barrow and R.M. Burstall.
Subgraph isomorphism, matching relational structures and maximal 
cliques.
\emph{Information Processing Letters} 4(4):83--84 (1976).

\item
R.M. Burstall.
Program proof, program transformation, program synthesis for
recursive programs.
Lecture notes for course on ``Data and program structures: syntax
and semantics''.
International School on Theory and Application of Computers, Erice,
Sicily (1976).

\item
R.M. Burstall and J. Darlington.
A transformation system for developing recursive programs.
\emph{Journal of the Association for Computing Machinery} 24(1):44--67
(1977).
Based on [\ref{reliable}].

\item \label{clear-theories}
R. Burstall and J. Goguen.
Putting theories together to make specifications.
Invited paper, \emph{Proc.\ 5th Intl.\ Joint Conf.\ on
  Artificial Intelligence}, Cambridge, Massachusetts, 1045--1058 (1977).

\item
R.M. Burstall. \label{NPL}
Design considerations for a functional programming language.
Invited paper, \emph{Proc.\ Infotech State of the Art Conf.\
  ``The Software Revolution''}, Copenhagen, 45--57 (1977).

\item Alan Bundy, Rod Burstall, Sylvia Weir \& Richard Young. 
\emph{Artificial Intelligence: An introductory course}. Edinburgh University Press (1978).

\item
R.M. Burstall and M. Feather.
Program development by transformations: an overview.
\emph{Les fondements de la programmation.
   Proc.\ Toulouse CREST Course on Programming}
(eds.\ M. Amirchahy and D. Neel).
Le Chesnay: IRIA-SEFI (1978).

\item
R.M. Burstall.
Concepts versus code in program design.
\emph{Proc.\ Euro IFIP '79}, London.
North-Holland (1979).

\item
R.M. Burstall and J.A. Goguen. \label{clear}
The semantics of CLEAR, a specification language.
Proc.\ 1979 Winter School on Abstract Software
Specifications, Copenhagen.
Springer Lecture Notes in Computer Science, Vol.\ 86, 292--332 (1980).

\item \label{CAT}
J.A. Goguen and R.M. Burstall.
CAT, a system for the structured elaboration of correct programs
 from structured specifications.
Technical Report CSL-118, SRI International, Menlo Park (1980).

\item
J.L. Weiner and R.M. Burstall.
Making programs more readable.
\emph{Proc.\ 4th Intl.\ Symp.\ on Programming}, Paris.
Springer Lecture Notes in Computer Science, Vol.\ 83, 372--341 (1980).

\item
R.M. Burstall.
Electronic category theory.
Invited paper, \emph{Proc.\ 9th Intl.\ Symp. on
  Mathematical Foundations of Computer Science}, Rydzyna.
Springer Lecture Notes in Computer Science, Vol.\ 88, 22--39 (1980).

\item \label{Hope}
R.M. Burstall, D.B. MacQueen and D.T. Sannella.
HOPE: an experimental applicative language.
\emph{Proc.\ 1980 LISP Conference}, Stanford, 136--143 (1980).

\item
R.M. Burstall and J.A. Goguen.
An informal introduction to CLEAR, a specification language.
\emph{The Correctness Problem in Computer Science} (eds. R. Boyer
and J. Moore).
Academic Press 185--213 (1981).
Republished in: \emph{Software Specification Techniques}
(eds. N. Gehani and A.D. McGettrick).
Addison Wesley, 363--389, 1986.

\item
R.M. Burstall and J.A. Goguen.
Algebras, theories and freeness:  an introduction for computer 
scientists.
\emph{Theoretical Foundations of Programming Methodology:
Lecture Notes of an International Summer School}, Marktoberdorf, 1981
(eds.\ M. Broy and G. Schmidt).
D. Reidel, 329--348 (1982).

\item
A. Pettorossi and R.M. Burstall.
Deriving very efficient algorithms for evaluating linear
recurrence relations using the program transformation technique.
\emph{Acta Informatica} 18:181--206 (1982).

\item
R.M. Burstall and N. Suzuki.
Sakura: a VLSI modelling language.
\emph{Proc.\ Conf.\ on Advanced Research in VLSI},
Cambridge, Massachusetts (1982).

\item
D.T. Sannella and R.M. Burstall.
Structured theories in LCF.
\emph{Proc.\ 8th Colloq.\ on Trees in Algebra and Programming},
L'Aquila.
Springer Lecture Notes in Computer Science, Vol.\ 159,
377--391 (1983).

\item \label{intro-institutions}
J.A. Goguen and R.M. Burstall.
Introducing institutions.
\emph{Proc. of Logics of Programming Workshop},
Pittsburgh.
Springer Lecture Notes in Computer Science, Vol.\ 164,
221--256 (1983).

\item
J.A. Goguen and R.M. Burstall.
Some fundamental algebraic tools for the semantics of computation.
Part 1: Comma categories, colimits, signatures and theories.
\emph{Theoretical Computer Science} 31(1--2):175--209 (1984).
Part 2: Signed and abstract theories.
\emph{Theoretical Computer Science} 31(3):263--295 (1984).

\item \label{pebble}
R.M. Burstall and B. Lampson.
A kernel language for abstract data types and modules.
Invited paper, \emph{Proc.\ Intl.\ Symp.\ on Semantics of
  Data Types}, Sophia-Antipolis.
Springer Lecture Notes in Computer Science, Vol.\ 173, 1--50
(1984).

\item
R.M. Burstall.
Programming with modules as typed functional programming.
Invited paper, \emph{Proc.\ Intl.\ Conf.\ on 5th Generation
  Computer Systems}, Tokyo, 103--112 (1984).

\item
D.E. Rydeheard and R.M. Burstall.
Monads and theories: a survey for computation.
Seminar on the Application of Algebra to Language Definition and
Compilation, Fontainebleau, 1982.
\emph{Algebraic Methods in Semantics} (eds. M. Nivat and
J.C. Reynolds).
Cambridge University Press, 575--605 (1985).

\item \label{tapsoft}
R.M. Burstall.
Inductively defined functions.
Invited paper, \emph{Proc.\ Intl.\ Joint Conf.\ on Theory
  and Practice of Software Development}, Berlin.
Springer Lecture Notes in Computer Science, Vol.\ 185, 92--96 (1985).

\item
J.A. Goguen and R.M. Burstall.
A study in the foundations of programming methodology:
specifications, institutions, charters and parchments.
\emph{Proc.\ Summer Workshop on Category Theory and Computer
  Programming}, Guildford, 1985.
Springer Lecture Notes in Computer Science, Vol.\ 240,
313--333 (1986).


\item
D.E. Rydeheard and R.M. Burstall.
A categorical unification algorithm.
\emph{Proc.\ Summer Workshop on Category Theory and Computer
  Programming}, Guildford, 1985.
Springer Lecture Notes in Computer Science, Vol.\ 240,
493--505 (1986).

\item
R.M. Burstall and D.E. Rydeheard.
Computing with categories.
\emph{Proc.\ Summer Workshop on Category Theory and Computer
  Programming}, Guildford, 1985.
Springer Lecture Notes in Computer Science, Vol.\ 240,
506--519 (1986).

\item
R.M. Burstall.
Inductively defined functions in functional programming
languages.
\emph{Journal of Computer and System Sciences} 34:409--421
(1987).
Based on [\ref{tapsoft}].

\item
R.M. Burstall.
Research in interactive theorem proving at Edinburgh
University.
Invited paper, \emph{Proc.\ 20th IBM Computer Science
  Symposium}, Shizuoka, Japan (1987).

\item \label{CCT}
D.E. Rydeheard and R.M. Burstall.
\emph{Computational Category Theory}.
Prentice-Hall (1988).

\item
B. Lampson and R.M. Burstall.
Pebble, a kernel language for modules and abstract data
types.
\emph{Information and Computation} 76:278--346 (1988).
Based on [\ref{pebble}].

\item
R.M. Burstall and F. Honsell.
A natural deduction treatment of operational semantics.
Invited paper, \emph{Proc.\ 8th Conf.\ on Foundations of
  Software Technology and Theoretical Computer Science}, Pune.
Springer Lecture Notes in Computer Science, Vol.\ 338,
250--269 (1988).


\item
R.M. Burstall. \label{Lego}
Computer assisted proof for mathematics: an introduction,
using the LEGO proof system.
\emph{Proc.\ IAM Conference on The Revolution in Mathematics Caused
  by Computing}, Brighton (1990).

\item
A. Tarlecki, R.M. Burstall and J.A. Goguen.
Some fundamental algebraic tools for the semantics of computation.
Part 3: Indexed categories.
\emph{Theoretical Computer Science} 91:239--264 (1991).

\item \label{reality}
R.M. Burstall.
Computing: yet another reality construction.
Invited paper, \emph{Software Development and Reality Construction}
(eds.\ C. Floyd, H. Z\"ullighoven, R. Budde and R. Keil-Slawik).
Springer Verlag (1992).

\item \label{institutions}
J.A. Goguen and R.M. Burstall.
Institutions: abstract model theory for specification and
programming.
\emph{Journal of the Association for Computing Machinery}
39(1):95--146 (1992).

\item \label{deliverables}
J. McKinna and R.M. Burstall.
Deliverables: a categorical approach to program development
in type theory.
Invited paper, \emph{Proc.\ 18th Intl.\ Symp.\ on
  Mathematical Foundations of Computer Science}, Gdansk.
Springer Lecture Notes in Computer Science, Vol.\ 711,
32--67 (1993).

\item
R.M. Burstall and R. Diaconescu.
Hiding and behaviour: an institutional approach.
\emph{A Classical Mind: Essays in Honour of C.A.R. Hoare}.
Prentice-Hall, 75--92 (1994).

\item
R.M. Burstall.
Terms, proofs and refinement.
Invited paper,
\emph{Proc.\ 9th IEEE Symp.\ on Logic in Computer Science},
Paris, 2--7 (1994).

\item
M. Sato, T. Sakurai and R.M. Burstall.
Explicit environments.
\emph{Proc.\ 4th Intl.\ Conf.\ on Typed Lambda Calculi and
  Applications}, L'Aquila.
Springer Lecture Notes in Computer Science, Vol.\ 1581,
340--354 (1999).

\item
R.M. Burstall. \label{PE}
ProveEasy: helping people learn to do proofs.
Invited paper, \emph{Computing: the Australasian Theory
  Symposium (CATS 2000)}, Canberra.
\emph{Electronic Notes in Theoretical Computer Science} 31 (2000).

\item \label{strachey}
R.M. Burstall.
Christopher Strachey --- understanding programming languages.
\emph{Higher-Order and Symbolic Computation} 13:51--55 (2000).

\item
H. Goguen, R. Brooksby and R.M. Burstall.
Memory management: an abstract formulation of incremental tracing.
\emph{Types for Proofs and Programs: Intl.\ Workshop TYPES'99 ---
  Selected Papers}.
Springer Lecture Notes in Computer Science, Vol.\ 1956,
148--161 (2000).

\item 	M.\ Sato, T.\ Sakurai, R.~M.~Burstall,
Explicit Environments. \emph{Fundam.\ Informaticae} 45(1--2): 79--115, (2001).

\item R.~M.~Burstall,
My Friend Joseph Goguen. 
\emph{Algebra, Meaning, and Computation, Essays Dedicated to Joseph A.\ Goguen on the Occasion of His 65th Birthday}
 (eds.\ Kokichi Futatsugi, Jean-Pierre Jouannaud, and Jos\'{e} Meseguer). Lecture Notes in Computer Science 4060: 25--30, Springer, (2006),

\end{enumerate}
\end{document}